\documentclass[aps,prd,twocolumn,groupedaddress,nofootinbib,showpacs,eqsecnum]{revtex4}

\usepackage{euscript,graphicx,amssymb,subfigure}

\usepackage{amsfonts}
\usepackage{amsmath}
\usepackage{amssymb}
\usepackage{graphicx}
\usepackage{euscript}
\usepackage{color}
\usepackage{subfigure}

\newcommand{\goodgap}{\hspace{\subfigtopskip} \hspace{\subfigbottomskip}}

\begin{document}

\title{Tachyon Field in Intermediate Inflation on the Brane}

\author{H. Farajollahi$^{1}$
\footnote{{\tt hosseinf@guilan.ac.ir}},
A. Ravanpak$^{1}$
\footnote{{\tt aravanpak@guilan.ac.ir}} }

\affiliation{$^1$Department of Physics, University of Guilan, Rasht, Iran}

\begin{abstract}

We propose a new model of inflation in the framework of brane cosmology in the presence of tachyonic field. we derive the
expressions for the model parameters in brane inflation, and estimate the observable parameters numerically
and find them to fit with the observational data.

\end{abstract}


\date{\today}
\maketitle

\section{Introduction}

Due to recent theoretical developments and accurate astronomical data, cosmology presents explosive activity which allows to use astrophysics to perform tests of fundamental theories, otherwise inaccessible to terrestrial accelerators. The universe evolution during its different periods leads to considering at some stage in the early universe an inflationary phase that is one of the most compelling solution to many long-standing problems of the standard hot big bang scenario, for example, the flatness, the horizon, and the monopole problems, among others \cite{Guth},\cite{Albrecht}. Besides, the inflation era has provided a causal interpretation of the origin of the observed anisotropy of the cosmic microwave background (CMB) radiation, and also the distribution of large scale structures \cite{Dunkley},\cite{Hinshaw}.

In higher dimensional cosmologies, implications of string/M-theory to Friedmann-Robertson-Walker (FRW) models in particular those related to brane cosmological models have recently attracted great deal of attention \cite{Sen}. In these models the
standard model of particle is confined to the brane, while gravitations propagate in the bulk spacetimes. The effect of the extra dimension induces additional terms in the Friedmann equation in the brane \cite{Binetruy}-\cite{Shir}. Specially, a quadratic term in the energy density generally makes it easier to obtain inflation in the early universe \cite{Maar}-\cite{Mohapatra2}.

Among inflationary models the particular model of intermediate inflation is of special interest. In this model expansion of the universe is slower than standard de Sitter inflation $(a=\exp(Ht))$, still faster than power law inflation $(a=t^p; p > 1)$ \cite{Barrow}-\cite{Rendall}. The model which was originally introduced as an exact inflationary solution for a particular scalar field potential, then become the slow-roll solution to potentials which are asymptotically of inverse power-law type, $V(\phi)\propto\phi^{-\beta}$ commonly used in quintessence models \cite{Ratra}-\cite{Liddle}. Nothing that with this type of potential inflation is everlasting and a mechanism has to be introduced to bring inflation to an end \cite{del}-\cite{Faraj}. The string theory which is the motivation behind intermediate inflationary model \cite{Sanyal} suggest that in order to have a ghost-free action high order curvature invariant corrections to the Einstein-Hilbert action must be proportional to the Gauss-Bonnet (GB) term \cite{Boul},\cite{Boul2}. The GB term as the inverse string tension, arise naturally as the leading order of the expansion to the low-energy string effective action \cite{Kolvis},\cite{Kolvis2}. These theories can be applied to solve the universe initial singularity, and explain the late time universe acceleration, among others \cite{Antoni}-\cite{Gognola}. In particular, in four dimension, for a dark energy model, the coupling of a dynamical dilatonic scalar field with GB term leads to the intermediate form of the solutions for scale factor, where the universe starts evolving with a decelerated exponential expansion, for $A = 2/(\kappa n)$, $f=1/2$, with $\kappa^2=8\pi G$ and $n$ as a constant \cite{Sanyal}. Thus, the idea that intermediate inflation comes from an effective theory at low dimension of a more fundamental string theory is in itself very attractive.

Generally we assume an inflationary phase driven by the potential or vacuum energy of a scalar field, the inflaton, where its dynamics is determined by the Klein-Gordon action. However, there are other non-standard scalar field actions have been used in cosmology. In particular, in k-inflation \cite{Armenda} higher-order scalar kinetic terms in the action can, without the help of the potential, drive an inflationary evolution.

 The tachyon inflation is a specific model of k-inflation for the scalar field $\phi$, with a positive potential $V(\phi)$, a maximum at $\phi = 0$ and minima as $|\phi|\rightarrow\infty$ where $V\rightarrow0$. The cosmological application of the gravity-tachyon models, including slow-roll inflation have been investigated by many authors \cite{Gibbons}-\cite{Pad}. Infact, many potentials with the above properties can drive inflation, which typically takes place at a scale characterized by the brane tension, $H\sim\lambda^{1/2}/M_{Pl}$, where $M_{Pl}=(8\pi G)^{-1/2}$. In addition, the rolling tachyon can contribute to the energy density of the universe with dust-like equation of state \cite{Sen},\cite{Sen2}. Thus, the tachyon may at the same time drive inflation and later behave as dark matter.

 So far, with the presence of tachyon field,  intermediate inflation in standard gravity has been studied by the authors in \cite{Campo}. Separately, also the standard scalar field in intermediate inflation both in standard gravity and brane cosmology has been investigated in \cite{Barrow3},\cite{Campo2}. In this paper an intermediate inflationary universe model driven by a tachyon scalar field in a brane world effective theory will be presented. The motivation for studying the tachyon inflation in the brane world scenario has the interesting feature that the inflaton potential comes
from higher dimensional gravity, or more generally, from bulk modes or string theory \cite{Shiu}. The outline of the paper is as follows. The next section presents a short review of the tachyonic-brane-intermediate inflationary phase. In section III we study the scalar and tensor perturbations
in our model and derive the cosmological parameters. With the recent observational data from WMAP we also constraint the parameters. Finally, in Section IV we present a conclusion.

\section{The tachyonic Brane Intermediate Inflationary Phase}

In this section, we begin with the modified Friedmann equation in five dimensional brane cosmology as
\begin{equation}\label{modfried}
H^2 =
\kappa\rho_\phi[1+\frac{\rho_\phi}{2\lambda}]+\frac{\Lambda_4}{3}+\frac{\xi}{a^4}\cdot
\end{equation}
where $H$ denotes the Hubble parameter, $\rho_\phi$ represents the
energy density of the matter field confined to the brane in form of a scalar field, $\kappa = 8\pi G/3 =
8\pi/(3m_p^2)$, $\Lambda_4$ is the four-dimensional cosmological
constant and the last term represents the influence of the bulk
gravitons on the brane, where $\xi$ is an integration constant. The
brane tension $\lambda$ relates the four and five-dimensional Planck
masses via the expression $m_p = \sqrt{3M_5^6/(4\pi\lambda)}$.

In the following we assume the high energy regime where $\rho_\phi \gg \lambda$. Also we
will take that the four dimensional cosmological constant to be
vanished and once inflation begins, the last term will rapidly
become unimportant, leaving us with the effective equation
\begin{equation}\label{highfried}
H^2 \simeq \beta\rho_\phi^2,
\end{equation}
where $\beta=\kappa^2/(2\lambda)$.

We also assume that the scalar field is confined to the brane, so that
its field equation has the standard form
\begin{equation}\label{conservation}
\dot\rho_\phi+3H(\rho_\phi+p_\phi)=0,
\end{equation}
where the dot means derivatives with respect to the cosmological time.
For a tachyonic scalar field, $\phi(t), $the energy density and the pressure are respectively given by
\begin{eqnarray}\label{tachyon}
  \rho_\phi=\frac{V(\phi)}{\sqrt{1-\dot\phi^2}}& , &p_\phi=-V(\phi)\sqrt{1-\dot\phi^2},
\end{eqnarray}
where $V(\phi)$ is the tachyonic scalar potential. From Eqs. (\ref{conservation}) and (\ref{tachyon}) the tachyonic scalar field equation of motion becomes
\begin{equation}\label{eomtachyon}
    \frac{\ddot\phi}{1-\dot\phi^2}+3H\dot\phi=-\frac{V^{'}}{V},
\end{equation}
where $V' = \partial V(\phi)/\partial\phi$. Also, with regards to the Eqs. (\ref{highfried}), (\ref{conservation}) and (\ref{tachyon}), one reads
\begin{equation}\label{velocitytach}
    \dot\phi=\sqrt{-\frac{\dot H}{3H^2}} \cdot
\end{equation}
For intermediate inflationary universe, from Eq. (\ref{velocitytach}) we derive exact solution for the tachyonic scalar field as
\begin{equation}\label{phi}
\phi=\phi_0+[\frac{4(1-f)}{3Af(2-f)^2}]^{1/2}t^{(2-f)/2},
\end{equation}
By using Eqs. (\ref{highfried}), (\ref{tachyon}), (\ref{velocitytach}) and (\ref{phi}) we find
\begin{equation}\label{potential}
V(\phi)=\alpha(\phi-\phi_0)^{2(f-1)/(2-f)}\sqrt{1-B(\phi-\phi_0)^{-2f/(2-f)}},
\end{equation}
where
\begin{equation}\label{alpha}
    \alpha=\beta^{-1/2}[Af(\frac{3(2-f)^2}{4(1-f)})^{(f-1)}]^{1/(2-f)},
\end{equation}
and
\begin{equation}\label{B}
    B=(\frac{1-f}{3Af})^{2/(2-f)}(\frac{(2-f)^2}{4})^{-f/(2-f)},
\end{equation}
From Eq. (\ref{phi}), one can find the the Hubble parameter as a function of $\phi$
\begin{equation}\label{hubble}
H(\phi)=\alpha\beta^{1/2}(\phi-\phi_0)^{2(f-1)/(2-f)},
\end{equation}
where without loss of generality we can assume $\phi_0=0$ .

The energy density for the tachyon field in the inflation era is of the
order of the potential, i.e. $\rho_\phi\sim V$. with slow-roll conditions, i.e. $\dot{\phi}^2 \ll 1$ and
$\ddot{\phi} \ll 3H\dot{\phi}$, Eqs. (\ref{highfried}) and (\ref{eomtachyon}) reduce to
\begin{equation}\label{highfriedred}
    H^2\approx\beta V^2,
\end{equation}
and
\begin{equation}\label{eomtachyonred}
    \frac{V^{'}}{V}\approx-3H\dot\phi,
\end{equation}
respectively. In addition, the scalar field potential, $V(\phi)$ becomes
\begin{equation}\label{potentialred}
    V(\phi)=\alpha\phi^{-\gamma},
\end{equation}
where
\begin{equation}\label{gamma}
    \gamma=\frac{2(1-f)}{2-f}.
\end{equation}
Note that this kind of potential does not present a minimum and thus is appropriate for investigation of
 reheating of the Universe in a nonstandard way.

The dimensionless slow-roll parameters, in terms of the model parameters can be written as
\begin{equation}\label{epsilon}
\varepsilon=-\frac{\dot H}{H^2}=\frac{{V^{'}}^2}{3\beta V^4}=\frac{\gamma^2}{3\beta\alpha^2}\phi^{2(\gamma-1)},
\end{equation}
\begin{equation}\label{eta}
\eta=-\frac{\ddot \phi}{H\dot \phi}=\frac{V^{''}}{3\beta V^3}-\frac{2{V^{'}}^2}{3\beta V^4}=\frac{\gamma(1-\gamma)}{3\beta\alpha^2}\phi^{2(\gamma-1)}.
\end{equation}
The number of e-folds between two different values $\phi(t = t_1) = \phi_1$ and
$\phi(t = t_2) = \phi_2 > \phi_1$ can also be expressed as
\begin{widetext}
\begin{equation}\label{efolds}
N = \int_{t_1}^{t_2} H dt = \int_{\phi_1}^{\phi_2} H \frac{dt}{d\phi}d\phi = \frac{3\beta\alpha^2}{2\gamma(1-\gamma)}({\phi_2}^{-2(\gamma-1)}-{\phi_1}^{-2(\gamma-1)}).
\end{equation}
\end{widetext}
From the form of the potential and with regards to the conditions for the inflation to occur, we can find $\phi_1$, at the beginning of inflation
\begin{equation}\label{phibegin}
    \phi_1=(\frac{3\beta\alpha^2}{\gamma^2})^{1/(2(\gamma-1))}.
\end{equation}
when $\varepsilon=1$.

\section{perturbation}

In this section, we explore the linear perturbation theory
in inflation, which includes both scalar and tensor perturbation. We use a linearly perturbed flat FRW of the form,
\begin{widetext}
\begin{equation}\label{p-metric}
    ds^2 = -(1+2C)dt^2+2a(t)D_{,i}dx^idt+a(t)^2[(1-2\psi)\delta_{ij}+2E_{,i,j}+2h_{ij}]dx^idx^j,
\end{equation}
\end{widetext}
where $C$, $D$, $\psi$ and $E$ are the scalar metric perturbation and $h_{ij}$ is the transverse-traceless tensor perturbation. By considering the local
conservation of the energy-momentum tensor, the conservation of the curvature perturbation, $\cal R$, holds for adiabatic
perturbations irrespective of the form of gravitational equations. Note that even though
the effect of bulk to the cosmological perturbations can not be trivially negligible, It can
be shown that the main correction of the spectrum in the brane-world inflation is just the
modification of the slow-roll parameters \cite{Koyama}. For a tachyon field the power
spectrum of the curvature perturbations is given by ${\cal P}_{\cal R}=(\frac{H^2}{2\pi\dot\phi})^2\frac{1}{Z_s}$ where $Z_s=V(1-\dot\phi^2)^{-3/2}$ \cite{Hwang}. Under slow-roll approximation, using Eqs. (\ref{highfriedred}) and (\ref{eomtachyonred}), the approximated power spectrum of the curvature perturbations is
\begin{equation}\label{perturbation}
{\cal P}_{\cal R}\approx(\frac{H^2}{2\pi\dot\phi})^2\frac{1}{V}=\frac{9\beta^3V^7}{4\pi^2{V^{'}}^2}.
\end{equation}
The scalar spectral index $n_s$ is given by $n_s-1 = d\ln{\cal P}_{\cal R}/d\ln k$, where the interval in wave number is
related to the number of e-folds by the relation $d\ln k(\phi) =
dN(\phi)$ \cite{Campo}. With attention to definition of ${\cal P}_{\cal R}$ and
also $dN=Hdt$, we arrive at the relation:
\begin{equation}\label{ns}
    n_s\approx1 - 3\varepsilon + 2\eta,
\end{equation}
or equivalently,
\begin{equation}\label{nsphi}
n_s\approx1-\frac{(5\gamma-2)\gamma}{3\beta\alpha^2}\phi^{2(\gamma-1)}.
\end{equation}
in terms of the model parameters which Eq. (\ref{ns}) is consistent to what derived in \cite{Calcagni} for a tachyon scalar field.

We clearly see that the Harrison-Zel'dovich model, i.e., $n_s=1$
occurs for $\gamma=2/5$ or from Eq. (\ref{gamma}), $f=3/4$. Obviously, for $\gamma \lessgtr 2/5$ (or equivalently $f\gtrless3/4$) we have $n_s\gtrless1$, respectively. The above results are similar to that obtained in \cite{Campo2} for a brane intermediate inflationary model with a standard scalar field.

Also, using (\ref{efolds}) and (\ref{phibegin}), we can re-express (\ref{nsphi}) in terms of the number of e-folding $N$, resulting in
\begin{equation}\label{ns-N1}
    n_s=1-\frac{(5\gamma-2)}{(2(1-\gamma)N+\gamma)}\cdot
\end{equation}
One of the interesting features of the seven-year data set from
Wilkinson Microwave Anisotropy Probe (WMAP) is that it hints at a
significant running in the scalar spectral index $n_{run} = dn_s/d\ln k$ \cite{Larson,Komatsu}. From (\ref{ns}) we get that the running of the scalar spectral
index becomes
\begin{equation}\label{nrun}
n_{run}=\frac{V}{V^{'}}(3\varepsilon^{'}-2\eta^{'})\varepsilon=\frac{2\gamma^2(1-\gamma)(2+\gamma)}{9\beta^2\alpha^4}\phi^{4(\gamma-1)}.
\end{equation}
where as already mentioned, primes denote differentiation with respect to $\phi$. In models with only scalar fluctuations the marginalized value for
the derivative of the spectral index is approximately -0.034 from
WMAP-seven year data only \cite{Larson,Komatsu}.

From (\ref{nsphi}) and (\ref{nrun}) we can write down a relation between spectral index and its running
\begin{equation}\label{nrun-ns}
    n_{run}=\frac{2(1-\gamma)(1-{n_s})^2}{(5\gamma-2)}.
\end{equation}
On the other hand, the generation of tensor perturbations during inflation would produce
gravitational waves. These perturbations in cosmology are more involved in our case, since
in brane-world gravitons propagate in the bulk. The amplitude of tensor perturbations is given by \cite{Campo2}
\begin{equation}\label{tensorperturb}
{\cal P}_g=24\kappa(\frac{H}{2\pi})^2 F^2(x),
\end{equation}
where $x = Hm_p\sqrt{3/(4\pi\lambda)} = 3Hm_p^2\beta^{1/2}/(4\pi)$ and
\begin{equation}\label{fx}
F(x)=[\sqrt{1+x^2}-x^2\sinh^{-1}(\frac{1}{x})]^{-\frac{1}{2}}.
\end{equation}
In our model (\ref{tensorperturb}) reduces to
\begin{equation}\label{tensorperturb2}
    {\cal P}_g=\frac{6\kappa\beta V^2}{\pi^2}F^2(x).
\end{equation}

From expressions (\ref{perturbation}) and (\ref{tensorperturb2}) we write the tensor-scalar ratio as
\begin{equation}\label{ratio}
r=\frac{{\cal P}_g}{{\cal P}_{\cal
R}}=\frac{8\kappa{V^{'}}^2}{3\beta^2V^5}F^2(x) \cdot
\end{equation}
By using expressions (\ref{potentialred}), (\ref{nsphi}) and (\ref{ratio}) we can obtain a relation between $n_s$ and $r$ as
\begin{widetext}
\begin{equation}\label{rns}
    r \approx
    (\frac{8\kappa}{3})\alpha^{1/(\gamma-1)}(\frac{\beta}{\gamma})^{(2-\gamma)/(2\gamma-2)}[\frac{3(1-n_s)}{5\gamma-2}]^{(3\gamma-2)/(2\gamma-2)}F^2(n_s).
\end{equation}
\end{widetext}
In \cite{Campo} the authors have shown that the tensor-scalar ratio, $r$, in terms of scalar spectral index $n_s$, for a tachyon field in standard intermediate inflationary model exactly coincides with the expression has been obtained for a standard scalar field. Interestingly, It can be shown that even in a brane intermediate inflationary scenario we get the same expression. In a similar manner we can rewrite Eq. (\ref{rns}), with attention to expressions (\ref{alpha}) and (\ref{gamma}) as
\begin{equation}\label{rnsrewrite}
    r \approx \frac{8\kappa(1-f)}{\beta^{1/2}}(Af)^{-1/f}[\frac{3-4f}{1-n_s}]^{(1-2f)/f}F^2(n_s).
\end{equation}
This equation has exactly the form obtained in \cite{Campo2}. Therefore, on the basic
of the brane intermediate inflation, the trajectories in the $r-n_s$ plane between standard field and
tachyon field can not be distinguished at lowest order. However, maybe tachyon inflation leads to a deviation at second order in
the consistency relations, like what has investigated in \cite{Campo} for a standard model.

From the potential obtained in (\ref{potentialred}) we can assume
that the tachyon is rolling from the top of the potential, i.e. $\varepsilon\ll\eta$ for
small $\phi$. In the following we will study the case in which the above condition imposes a constraint on $f$ \cite{Fair,Campo,Steer}. To see this we write down the ratio between $\eta$ and $\varepsilon$ as
\begin{equation}\label{etaepsilon}
    \mid\frac{\eta}{\varepsilon}\mid=\frac{1-\gamma}{\gamma},
\end{equation}
in which for $\eta>\varepsilon$, considering (\ref{gamma}) we obtain the constraint $f>2/3$, similar to the result found in \cite{Campo}.

\begin{figure*}
\centering
\subfigure{\includegraphics[width=6cm]{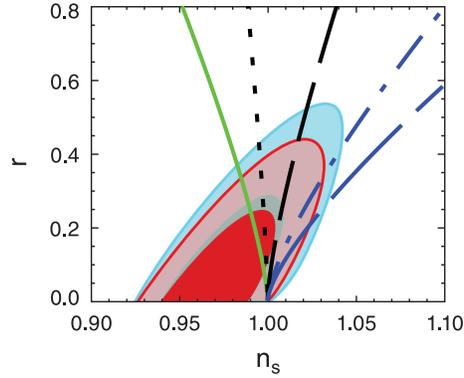}} \goodgap
\caption{The plot shows $n_s$ versus $r$ for our models and they are compared with the WMAP data
(five year and seven year). The green solid-line, black dot-line and blue dot-dash-line curves represent the tachyon lowest order case for different values of $f=0.1, 0.7$ and $0.9$, respectively. The black dash-line and blue long-dash-line curves specify the case $\varepsilon\ll\eta$ for $f=0.7$ and $0.9$, respectively
The two contours correspond to the 68\% and 95\% levels of confidence\cite{Larson}.}
\label{fig: f-r-ns}
\end{figure*}

The scalar spectral index $n_s$, for $\varepsilon\ll\eta$, is given by
\begin{equation}\label{nssimple}
    n_s\approx1+2\eta,
\end{equation}
where by using Eq.(\ref{eta}) becomes
\begin{equation}\label{nssimple2}
    n_s\approx1+\frac{2\gamma(1-\gamma)}{3\beta\alpha^2}\phi^{2(\gamma-1)}.
\end{equation}
At the beginning of inflation with the condition $|\eta|=1$, Eq.(\ref{nssimple2}) can be re-expressed in terms of the number of e-folding $N$ by
\begin{equation}\label{ns-N}
    n_s=1+\frac{2}{1+2N},
\end{equation}
where make it impossible to have Harrison-Zel'dovich spectrum. Using Eq. (\ref{ns-N}) we obtain the running scalar spectral index as
\begin{equation}\label{nrun-N}
    n_{run}=\frac{-4}{(1+2N)^2}.
\end{equation}
Also, from (\ref{ns-N}) and (\ref{nrun-N}), the relation between $n_s$ and $n_{run}$ becomes
\begin{equation}\label{ns-nrun}
    n_{run}=-(n_s-1)^2.
\end{equation}
On the other hand, from Eqs.(\ref{alpha}), (\ref{potentialred}), (\ref{gamma}), (\ref{ratio}) and (\ref{nssimple2}) we can obtain the tensor-scalar ratio as
\begin{equation}\label{r-ns-simple}
    r \approx \frac{8\kappa(1-f)}{\beta^{1/2}}(Af)^{-1/f}[\frac{-f}{1-n_s}]^{(1-2f)/f}F^2(n_s).
\end{equation}
In Fig.(\ref{fig: f-r-ns}) the tensor-scalar ratio, $r$, with respect to the spectral index, Eq. (\ref{rnsrewrite}), for different values of the parameter $f$ in the tachyon lowest order is shown in green solid-line, black dot-line and blue dot-dash-line. We compare it with the special case for $\varepsilon\ll\eta$, Eq. (\ref{r-ns-simple}),  (shown by the black dash-line and the blue long-dash-line)and with respect to the observational data. The two contours in the plot show the 68\% and 95\% levels of confidence, for the $r-n_s$ plane, which are defined at $k_0 = 0.002$ Mpc$^{-1}$. The seven-year WMAP data places stronger limits on $r$ (red, $r<0.36$) than five-year data (blue, $r<0.43$)\cite{Larson}.

\begin{figure*}
\centering
\subfigure{\includegraphics[width=6cm]{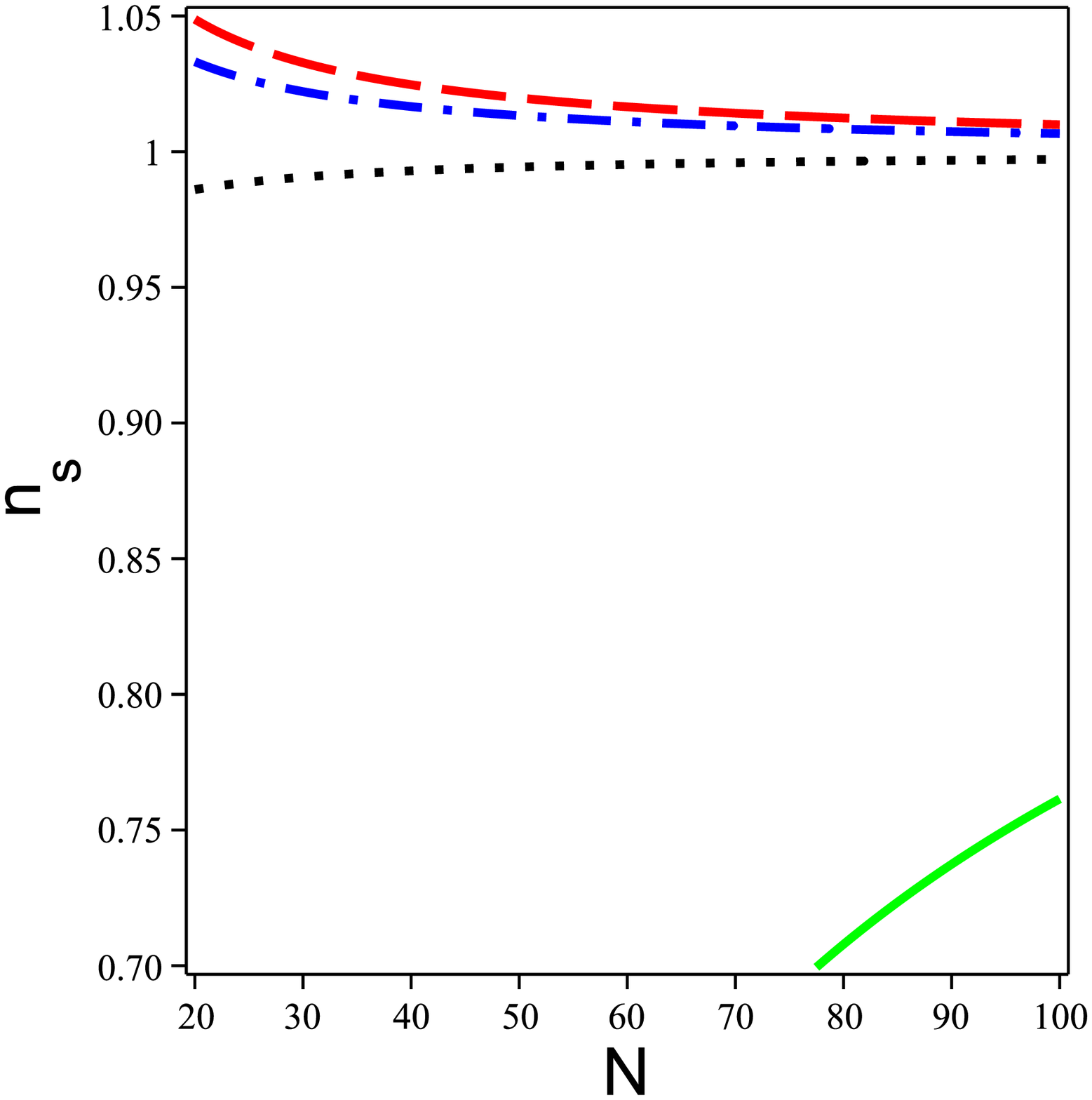}} \goodgap
\subfigure{\includegraphics[width=6cm]{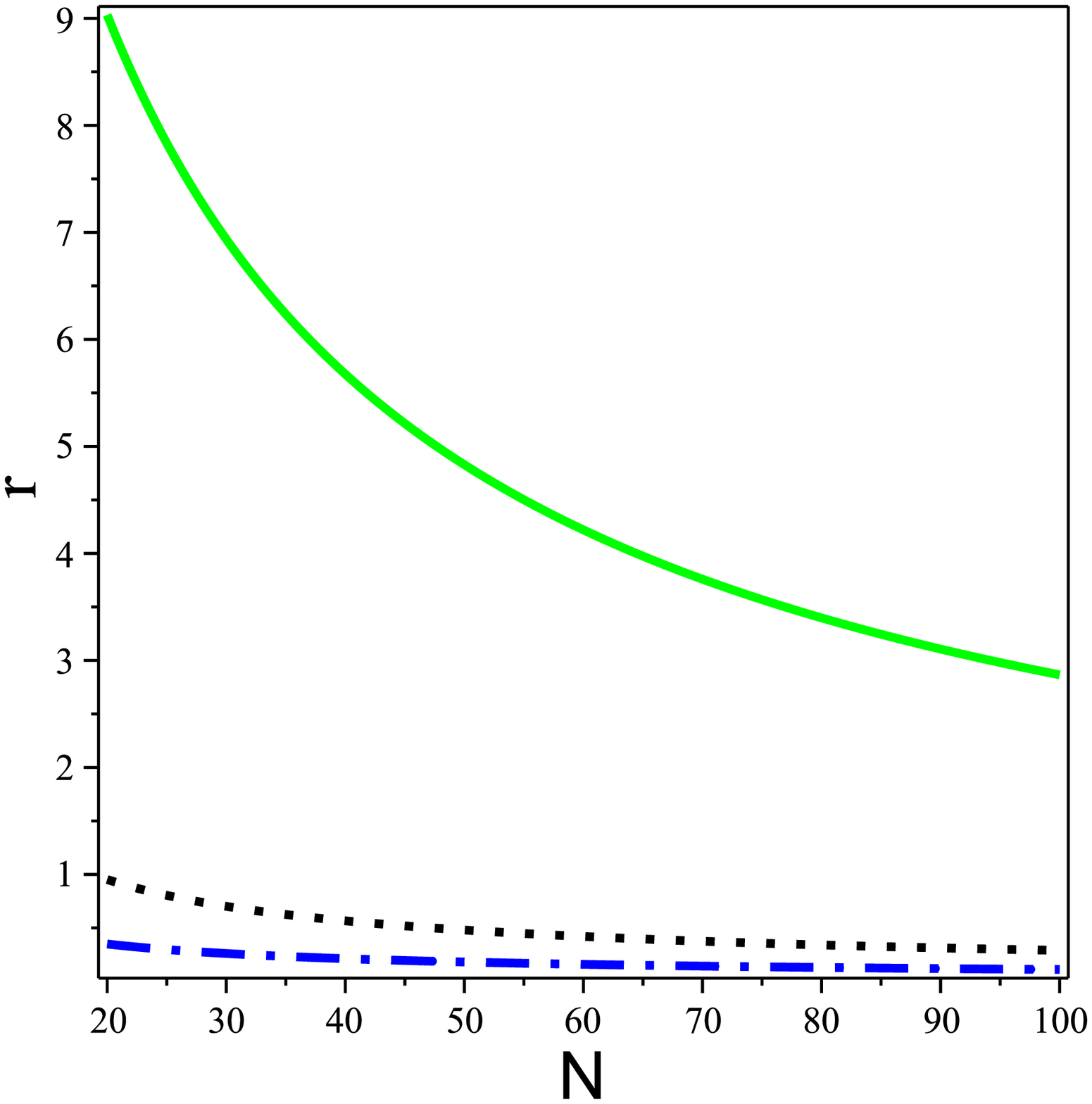}} \goodgap
\caption{The left plot shows $n_s$ versus $N$ for $f=0.1$(green solid-line), $f=0.7$(black dot-line) and $f=0.9$(blue dot-dash-line) in the tachyon lowest order case and also the red dash-line in the regime $\varepsilon\ll\eta$. The right plot shows $r$ versus $N$. The green solid-line, black dot-line and blue dot-dash-line are related to $f=0.1, 0.7$ and $0.9$, respectively, in the tachyon lowest order case. For the regime $\varepsilon\ll\eta$, the curves for $f=0.7$ and $f=0.9$ are exactly the same.}
\label{fig: f-r-N}
\end{figure*}

For tachyon lowest order case any value of the parameter $f$, (restricted to the range $0<f<1$), is well supported by the data, as can be seen from Fig.(\ref{fig: f-r-ns}). The curve $r=r(n_s)$ for WMAP 7-years, for $f=0.1$ enters the 95\% confidence region for $r\simeq0.33$, for $f=0.7$ enters the 95\% confidence region for $r\simeq0.37$ and for $f=0.9$ enters the 95\% confidence region for $r\simeq0.34$. These values by some numerical calculations correspond to $N\approx1758$, $N\simeq71$ and $N\simeq21$, respectively (See, Fig.(\ref{fig: f-r-N})(right panel)).

On the other hand, when we consider the regime where $\varepsilon\ll\eta$, any value of the parameter $f$ which satisfies the constraint $f>2/3$, is well supported by the data, as can be seen from Fig.(\ref{fig: f-r-ns}). We see that for $f=0.7$ the curve $r=r(n_s)$ for WMAP 7-years, enters the 95\% confidence region for $r\simeq0.43$ which corresponds to $N\simeq58$ and for $f=0.9$ enters the 95\% confidence region for $r\simeq0.21$ which corresponds to $N\simeq40$. One can check that the curves $r=r(N)$ in this case are exactly the same as ones in Fig.(\ref{fig: f-r-N})(right panel) for those values of $f$ that satisfy the condition $f>2/3$.

In Fig.(\ref{fig: f-r-N})(left panel) we have also shown the curves $n_s=n_s(N)$ for both models. We should notice that when we consider the regime $\varepsilon\ll\eta$, there is only one curve ( the red dash-line) that does not depend explicitly on $f$ parameter.

\begin{figure*}
\centering
\subfigure{\includegraphics[width=6cm]{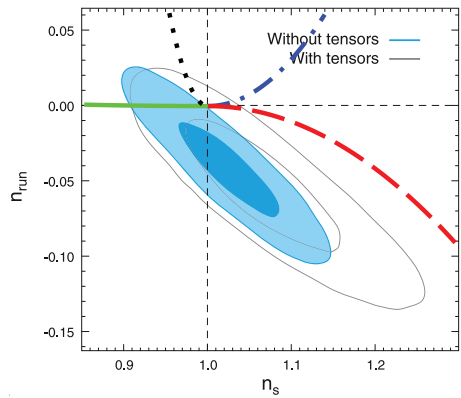}} \goodgap
\caption{The plot shows $n_{run}$ versus $n_s$ for $f=0.1$(green solid-line), $f=0.7$(black dot-line) and $f=0.9$(blue dot-dash-line) in the tachyon lowest order case and also the red dash-line in the regime $\varepsilon\ll\eta$.}
\label{fig: f-nrun-ns}
\end{figure*}

In Fig.(\ref{fig: f-nrun-ns}) we represent the dependence of the running of the scalar spectral index on the spectral index for different values of the parameter $f$. In this plot and from \cite{Dunkley}, two-dimensional marginalized limits for the spectral index, $n_s$, defined at $k_0 = 0.002$ Mpc$^{-1}$ and the running of the index, $n_{run}$, in which models with or without tensor contribution marginalized over, are shown.

We observe numerically that with a tensor contribution the $n_{run}-n_s$ curve for tachyon lowest order case, for $f=0.7$ enters the 95\% confidence region for $n_{run}\simeq0.023$, for $f=0.9$ enters the 95\% confidence region for $n_{run}\simeq-0.003$ and for $f=0.1$ it is almost tangent to $n_{run}=0$. Also, for the case $\varepsilon\ll\eta$ the curve in $n_{run}-n_s$ plane enters the 95\% confidence region for $n_{run}\simeq-0.002$.

\section{Summary}

We have analyzed the intermediate
inflationary scenario in the brane  cosmology in the presence of tachyon field. An exact solution of the Friedmann equation on the brane for a flat Universe containing tachyonic scalar potential has been obtained. We have also found in the slow-roll approximation, the slow roll parameters in terms of the tachyon potential and its derivatives. Furthermore, explicit expressions for the corresponding power spectrum of the curvature
perturbations ${\cal P}_{\cal R}$, scalar spectrum index $n_s$ and its running $n_{run}$ and also tensor-scalar ratio $r$ are derived.

In order to find the constraints on our model parameters and compare them with the observational data, some numerical calculations for  $n_s$, $n_{run}$ and $r$  are performed. The results are in agreement for the tachyon lowest order case and the regime in which $\varepsilon\ll\eta$. The model for the tachyon lowest order case and in the special case of $\varepsilon\ll\eta$ is well supported by the observational data for any value of $f$ within the allowed range. In the tachyon lowest order case, we noted that the parameter $f$, lies in the range $0<f<1$ whereas in the $\varepsilon\ll\eta$ case, it further constrain to $2/3<f<1$.

The work in this paper is only one part of the whole framework of the inflation theory. The inclusion of the reheating epoch and transition to the standard cosmology can give us new constraints on the model parameters as well as scalar and tensor perturbation parameters.

\end{document}